\documentclass[journal=nature,a4paper,manuscript=article]{article}

\usepackage{graphicx,setspace}
\usepackage{caption}
\large
\usepackage{amsmath}
\usepackage{float}
\usepackage{upgreek}
\usepackage{color}
\usepackage{lineno}
\usepackage{soul}
\usepackage{url}
\usepackage[backend=biber, bibstyle=nature, citestyle=numeric-comp,
sorting=none, doi=false, isbn=false, url=false, eprint=false, maxbibnames=5]{biblatex}
\usepackage[left=2.5cm,right=2.5cm]{geometry}
\usepackage{epstopdf}

\usepackage[colorlinks = true,
linkcolor = blue,
urlcolor = blue,
citecolor = blue,
anchorcolor = blue]{hyperref}

\newcommand{\slfrac}[2]{\left.#1\middle/#2\right.}

\title{Spin-order-dependent magneto-elastic coupling in two dimensional antiferromagnetic MnPSe$_3$ observed through Raman spectroscopy}

\usepackage{authblk}

\author[1]{Daniel J. Gillard*}

\author[2]{Daniel Wolverson}

\author[1]{Oscar M. Hutchings}

\author[1]{Alexander I. Tartakovskii}

\affil[1]{Department of Physics and Astronomy, University of Sheffield, Sheffield S3 7RH, UK}
\affil[2]{Department of Physics, University of Bath, Bath, BA2 7AY, UK}

\affil[*]{Corresponding author: \texttt{d.j.gillard@sheffield.ac.uk}}

\begin{document}
	
	\maketitle
	
	\section*{Abstract}
	{Layered antiferromagnetic materials have emerged as a novel subset of the two-dimensional family providing a highly accessible regime with prospects for layer-number-dependent magnetism. Furthermore, transition metal phosphorous trichalcogenides, MPX3 (M = transition metal; X = chalcogen) provide a platform for investigating fundamental interactions between magnetic and lattice degrees of freedom providing new insights for developing fields of spintronics and magnonics. Here, we use a combination of temperature dependent Raman spectroscopy and density functional theory to explore magnetic-ordering-dependent interactions between the manganese spin degree of freedom and lattice vibrations of the non-magnetic sub-lattice via a Kramers-Anderson super-exchange pathway in both bulk, and few-layer, manganese phosphorous triselenide (MnPSe$_3$). We observe a nonlinear temperature dependent shift of phonon modes predominantly associated with the non-magnetic sub-lattice, revealing their non-trivial spin-phonon coupling below the N{\'e}el temperature at 74 K, allowing us to extract mode-specific spin-phonon coupling constants.
	}
	
	\section*{Introduction}
	
	Devices and structures constructed using two-dimensional (2D) layered materials\supercite{Geim2013} have garnered an increasing amount of research interest ever since the discovery of isolated monolayers of graphene,\supercite{Novoselov2004} transition metal dichalcogenides (TMDs),\supercite{Mak2010,Manzeli2017} and beyond.\supercite{Choi2017,Scarano2021} The number of different 2D materials encompassing and driving this area of research is rapidly growing.\supercite{Scarano2021} Most recently, the focus has turned to the discovery of intrinsic long range magnetic ordering in low dimensional van der Waals (vdW) crystals, offering of a highly accessible regime in which to study 2D magnetism.\supercite{Park2016a,Gong2017,Burch2018,Gibertini2019,Huang2017,Fei2018,Lee2016a,Kim2019,Lyons2020} Spin fluctuations are expected to be strongly enhanced in the low dimensional limit,\supercite{Huang2018,Samarth2017,Lado2017,Prosnikov2018,Kim2019a} while control and tuning of magnetic states and properties should be easily achieved through engineering perturbations within the system such as strain, alloying, light coupling, gating, proximity effects, and moir\'e patterns, in a similar manner to conventional 2D materials.\supercite{Manzeli2017,Burch2018,Gibertini2019}

	A key challenge however, is the detection of such long range magnetic ordering, along with the underlying domains and associated magnetic fluctuations and excitations. Standard techniques currently employed to detect such occurrences in bulk crystals, thin film and even nanoparticles suspended in solutions, such as superconducting quantum interference device (SQUID) magnetometry and neutron scattering, have limited applicability in 2D magnetic materials due to the reduced volumes.\supercite{Hellman2017} 
	
	Optical techniques can offer greater insights due to the increased spacial resolution available. Spectroscopic magnetic circular dichroism,\supercite{Ando1992,Gong2017,Huang2017,Burch2008} which takes advantage of the Faraday and Kerr effects, can provide an insight into the magnetic behaviour of a few layer sample, such as the strength and sign of the exchange, along with the identification of bands relevant to the magnetic ordering.\supercite{Ando1992,Burch2008} The first evidence of long range magnetic ordering in a single layer of vdW magnetic material was discovered in FePS$_3$ using Raman spectroscopy. \supercite{Wang2016a,Lee2016a} Raman spectroscopy\supercite{Raman1928,Wolverson2008} has successfully been used to reveal the magnetic exchange interaction strength via measurements of the two-magnon joint density of states,\supercite{Sandilands2010} as well as uncovering novel topology via measurement of fractional spin excitations.\supercite{Nasu2016} In this investigation, we employ high-resolution temperature-dependent Raman spectroscopy in order to probe the magnetic spin-ordering-dependent magneto-elastic interactions of bulk, and few layer, MnPSe$_3$.
	
	\begin{figure}[t!]
		\centering
		\includegraphics[width=0.9\textwidth]{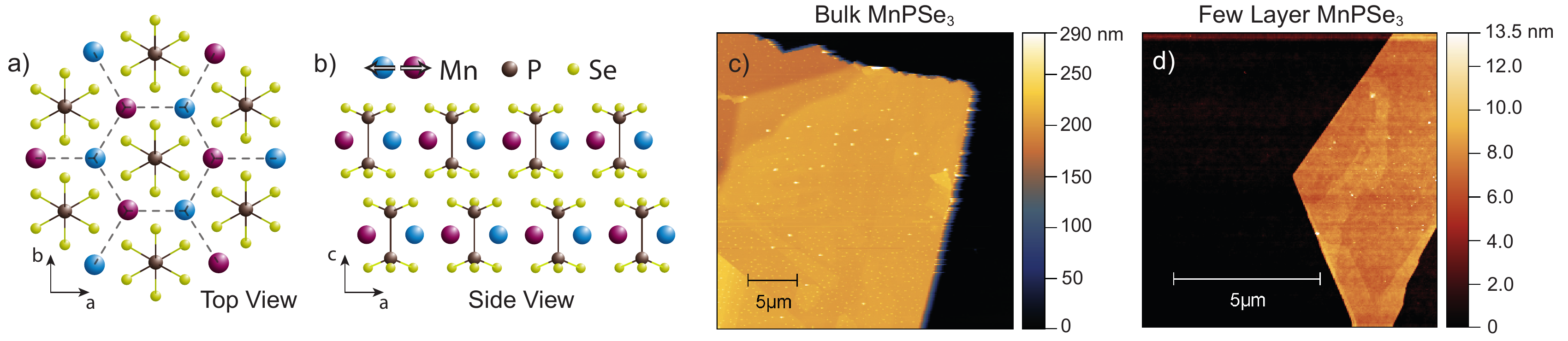}
		\caption{\label{Fig 1} {\bf Structure of MnPSe$_3$. a, b)} The MnPSe$_3$ crystal structure as viewed from the top (a) and the side (b). Manganese atoms are shown in purple and blue (depending on the spin state), while phosphorus is shown in brown, and selenium in yellow. Mn$^{2+}$ ions are arranged in a hexagonal lattice and form the antiferromagnetic magnetic structure, with the in -plane spin state highlighted by colour. The [P$_2$Se$_6$]$^{4-}$ ions possess no inherent magnetism and are arranged in a dumbbell-like formation centred within the Mn$^{2+}$ atoms. {\bf c, d)} Atomic force microscopy images of c) a bulk (209 nm), and d) a 9L (7.3 nm) MnPSe$_3$ sample. Lateral scale bars are 5 nm.}
	\end{figure}

	MnPSe$_3$ is a layered vdW magnetic material belonging to the transition metal phosphorus trichalcogenide, {\emph M}P{\emph X}$_3$ ({\emph M} = transition metal; {\emph X} = chalcogen), family with each layer possessing a $D_{3d}$ symmetry.\supercite{Grasso1999,Chittari2016,Wiedenmann1981a,Makimura1993,Ni2021a} The unit cell of MnPSe$_3$ can be considered as two distinct atomic groups of Mn$_2$ and P$_2$Se$_6$, formed as a dumbell-like structure of two PSe$_3$ tetrahedrons either end of a vertically (out-of-plane) orientated P-P bond, centred within the ionic bonded Mn hexagonal lattice (see Fig.\ref{Fig 1}a). Below the N{\'e}el temperature of 74 K, the Mn$^{2+}$ ions form inversion-breaking N\'eel type ordering with S=$\slfrac{5}{2}$ spins aligned in-plane (spin dimensionality of 1\supercite{Gibertini2019}) within an antiferromagnetic N{\'e}el type lattice, as shown in Fig.\ref{Fig 1}b.\supercite{Grasso1999,Chittari2016,Wiedenmann1981a,Makimura1993,Ni2021a} The P$_2$Se$_6$ cluster acts as an intermediary anion for a Kramers-Anderson super-exchange pathway between neighbouring magnetic Mn$^{2+}$ ions, formed through the spin state of the Mn$^{2+}$ donor electrons coupling to that of the receiving [P$_2$Se$_6$]$^{4-}$ anion.\supercite{Kramers1934,Anderson1950} This effectively couples the magnetic static spin-state correlation function of the manganese sub-lattice to the phonon modes of the phosphorus triselenide.\supercite{Lockwood1988,Ghosh2021,Casto2015,Fainstein2000,Prosnikov2018} 
	
	While 2D magnetic materials are being studied more actively, MnPSe$_3$ remains relatively unexplored. Recent investigations include evidence of phonon-magnon hybridisation, observed between a 2-magnon continuum and the low frequency Raman modes of heavy Mn ions,\supercite{Mai2021} and is also explored in this work. Using polarisation resolved second harmonic generation, it is possible to observe the different antiferromagnetic domains within a MnPSe$_3$ flake.\supercite{Ni2021a} The N\'eel vector in AFM domains is seen to reliably switch back and forth as a function of thermal cycle, along with a controllable rotation of the N\'eel vector via the application of uniaxial strain.\supercite{Ni2021a} The direct Mn-Mn nearest neighbour magnetic exchange interactions, both in-plane and out-of-plane, has been explored using a combination of DFT and neutron scattering techniques.\supercite{Calder2021} Theoretical investigations into the electronic and magnetic properties of MnPSe$_3$ predict a switching to ferromagnetic ordering when a large ($\sim$ 10$^{14}$ cm$^{-2}$) p- or n-doping is introduced to MnPSe$_3$.\supercite{Chittari2016} Further DFT calculations of MnPSe$_3$/MoS$_2$\supercite{Pei2017}, MnPSe$_3$/WS$_2$\supercite{Sharma2020} and MnPSe$_3$/CrSiTe$_3$\supercite{Pei2018} heterostructures, with an emphasis on strain tunability and spin-valley physics highlight the potential for MnPSe$_3$-based heterostructures with both TMDs and other 2D magnetic materials.
	
	In this study, a combination of temperature dependent low-frequency Raman spectroscopy and density functional theory (DFT) is used to probe the atomic vibrations of the non-magnetic ions. These phonons are used as a proxy by which to investigate the changing long range antiferromagnetic spin-ordering in MnPSe$_3$, via the aforementioned spin-phonon coupling, as the temperature is increased from 5 K through the N{\'e}el temperature (T$_N$) at 74 K, up to room temperature.\supercite{Lockwood1988,Ghosh2021,Casto2015,Fainstein2000,Prosnikov2018} We also consider the impact of sample thickness on spin-phonon coupling. 
		
	\section*{Results and Discussion}
	
	Flakes of MnPSe$_3$ are prepared via micromechanical exfoliation and thicknesses are determined via atomic force microscopy (AFM). The thinnest flakes measured have a thickness of 7.3 nm (9 layers\supercite{Ni2021a}) (Fig.\ref{Fig 1}d) and 7.9 nm (10 layers\supercite{Ni2021a}), while the thickest flake measured more than 200 nm (Fig.\ref{Fig 1}c), considered here as `bulk' material. Measurements are also taken on 14.1 nm ($\sim$ 18 layers) and 17.4 nm ($\sim$ 22 layers) thick MnPSe$_3$ flakes. It has been suggested that MnPSe$_3$ possesses a non-negligible interlayer magnetic coupling\supercite{Calder2021,Chittari2016} such that long range magnetic ordering, as observed in bulk samples at temperatures below the N{\'e}el transition, will be modified as we approach the 2D limit. The exact thickness at which the magnetic ordering might begin to weaken is currently unknown, however the results presented here suggest it is not above 9 layers.
	
\begin{figure}[t!]
	\centering
	\includegraphics[width=0.9\textwidth]{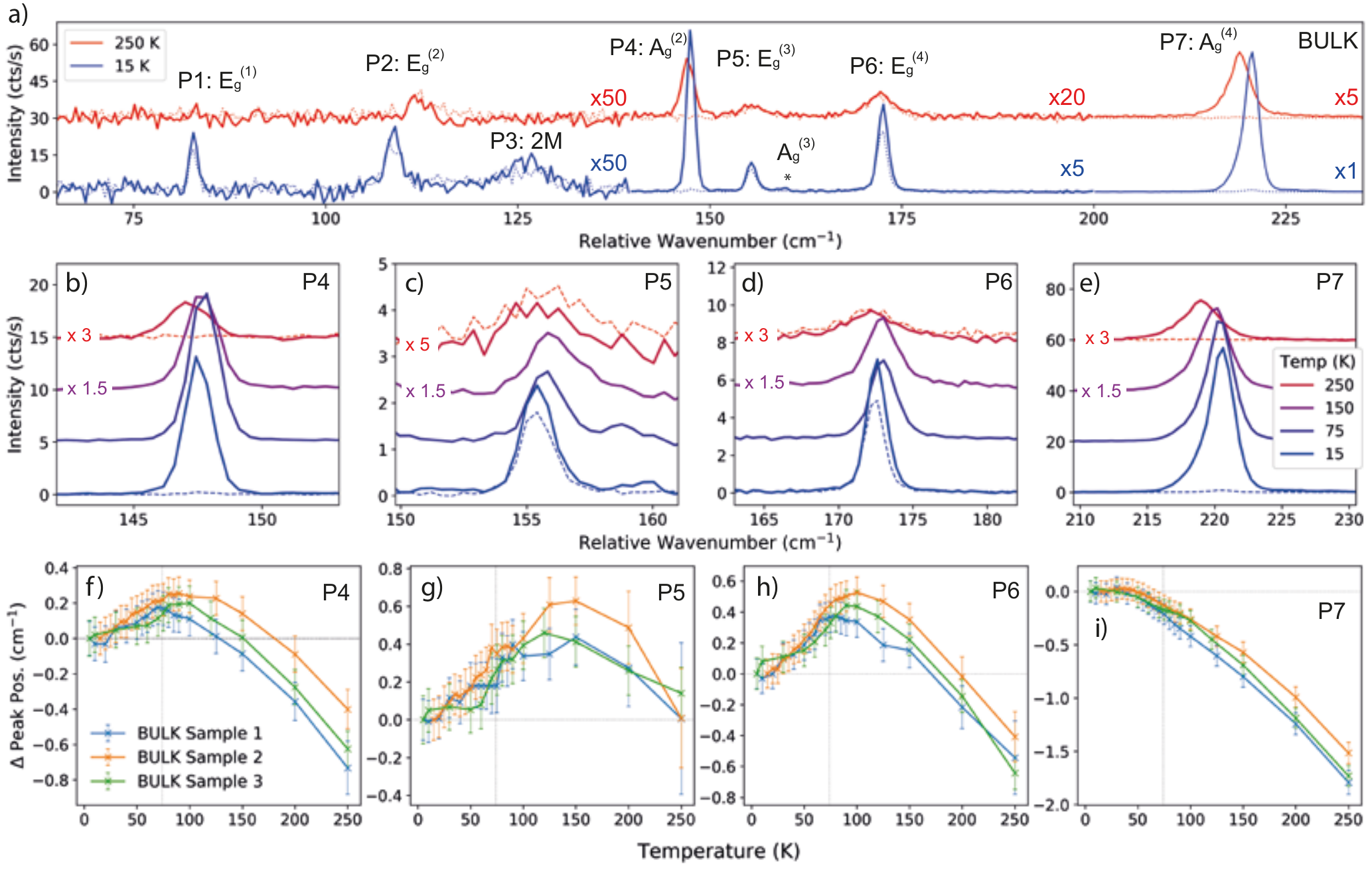}
	\caption{\label{Fig 2} {\bf Raman spectroscopy of bulk MnPSe$_3$. a)} Raman spectra of bulk MnPSe$_3$ at low (15 K), and high (250 K), displaying clearly identifiable peaks. {\bf b-e)} Temperature dependence of the non-magnetic vibrational modes P4 through to P7. A clear shift and change in intensity is observed as temperature is increased. {\bf f-i)} Change in peak position as a function of temperature determined from three separate bulk MnPSe$_3$ samples. The zero-line (Initial peak position at T=0 K) is highlighted with a horizontal dotted line. A vertical dotted line highlights the N{\'e}el temperature of 74 K. For visual purposes, some peak intensities have been magnified by the given factors.
	}
\end{figure}

	Initial observations of the low frequency Raman spectra of MnPSe$_3$ are carried out on three independent bulk samples, all showing a similar temperature dependence of the peak positions. The bulk MnPSe$_3$ Raman response at a temperature of 15 K and 250 K is shown in Figure \ref{Fig 2}a, and is consistent with previous investigations.\supercite{Mai2021,Calder2021,Makimura1993} The MnPSe$_3$ bulk Raman spectrum typically consists of seven main peaks, labelled here P1 at 84 cm$^{-1}$, through to P7 at 221 cm$^{-1}$. Two additional, but much dimmer, peaks can often be observed above the noise level at approximately 50 cm$^{-1}$ (not shown here), and 160 cm$^{-1}$ (labelled * in Fig.\ref{Fig 2}a).\supercite{Mai2021}
	
	Density functional theory is also used to corroborate the identity of these vibrational modes, providing further insight into the characteristics of each mode, which will be discussed in depth later. Each mode consists of contributions from one, or both, of the magnetic, Mn$^{2+}$, and non-magnetic, [P$_2$Se$_6$]$^{4-}$, sub-lattices. The two lowest frequency, immediately visible, Raman lines of P1 (84 cm$^{-1}$) and P2 (109 cm$^{-1}$), are largely comprised of in-plane lattice vibrations of the heavy manganese Mn$^{2+}$ ions, with smaller contributions from the non-magnetic phosphorus triselenide ions. Oppositely, the relatively unexplored P4 (148 cm$^{-1}$), P5 (156 cm$^{-1}$), P6 (173 cm$^{-1}$), and P7 (221 cm$^{-1}$) are formed mostly (or in the case of P4, entirely) due to vibrations of the non-magnetic ions, as detailed later in this study.
	
	The broad dim peak seen at 126 cm$^{-1}$, labelled in Figure \ref{Fig 2}a as P3, is identified as 2-magnon (2M) scattering, hybridising with P1 and P2. Consistent with previous reports,\supercite{Mai2021,Vaclavkova2020} the 2-magnon excitation (P3) is seen to red-shift as the temperature is increased and the magnetic ordering is destabilised with increasing thermal fluctuations. As P3 shifts through the two phonon modes, a broadening of the peak width, along with an associated drop in phonon lifetime, highlights the point of maximum hybridisation (See SI Note 4). The 2-magnon peak is a purely magnetic phenomenon, and therefore disappears when the magnetic ordering is lost at 74 K. As a consequence of such hybridisation, P1 and P2 present atypical temperature dependent behaviour and are seen to be blue-shifted at 250 K, when compared to 15 K, as can be seen in Figure \ref{Fig 2}a.
	
	In this investigation we concentrate on the previously unstudied phonon modes labelled here P4, P5, P6, and P7. Figure \ref{Fig 2}b-e details the Raman spectra of these four phonon modes at select temperatures of 15 K, 75 K, 150 K, and 250 K, highlighting the shift of each mode as a function of temperature. Solid lines show the Raman response detected in a co-polarised basis (HH), while the dashed line show the cross-polarised (HV) response, highlighting the polarisation dependence of each mode. We observe a near-unity polarisation degree $(I_{HH} - I_{HV})/(I_{HH} + I_{HV})$ of P4 (148 cm$^{-1}$) and P7 (221 cm$^{-1}$) that is maintained throughout the temperature range measured in each sample, while P5 (156 cm$^{-1}$) and P6 (173 cm$^{-1}$) are polarisation independent (observed in both HH and HV).\supercite{Liu2020c,Makimura1993,Mai2021} 
	
	Figure \ref{Fig 2}f-i shows the change of peak position of the four phonon modes due to the increasing temperature across three different bulk MnPSe$_3$ samples. The phonon modes are fitted using a Voigt curve, and peak position extracted at each temperature point, before subtracting the peak position at the lowest temperature. Each phonon line displays a distinct nonlinear temperature dependent behaviour that is repeated throughout multiple bulk samples. The most intense peak, P7 (A$_g^{(4)}$; 221 cm$^{-1}$), is seen to initially be constant with temperature increase until the N{\'e}el temperature is reached, at which point the peak is observed to start shifting towards lower frequencies. The other three peaks, P4 (A$_g^{(2)}$; 148 cm$^{-1}$), P5 (E$_g^{(3)}$; 156 cm$^{-1}$), and P6 (E$_g^{(4)}$; (173 cm$^{-1}$), initially tend towards higher frequencies until the N{\'e}el temperature is reached, when similarly to P7, the trend is inverted and the peak position is observed to shift to lower frequencies. P4 is seen to reach a maximum peak shift, $\sim$ 0.2 cm$^{-1}$, close to the N{\'e}el transition, while P5 and P6 both continue slightly beyond T$_N$ reaching $\sim$ 0.4 cm$^{-1}$. A specific change in trend observed around the N{\'e}el temperature is indicative of a magnetic ordering origin of this phenomena, as we discuss below.
	
	\begin{figure}[t!]
		\centering
		\includegraphics[width=0.9\textwidth]{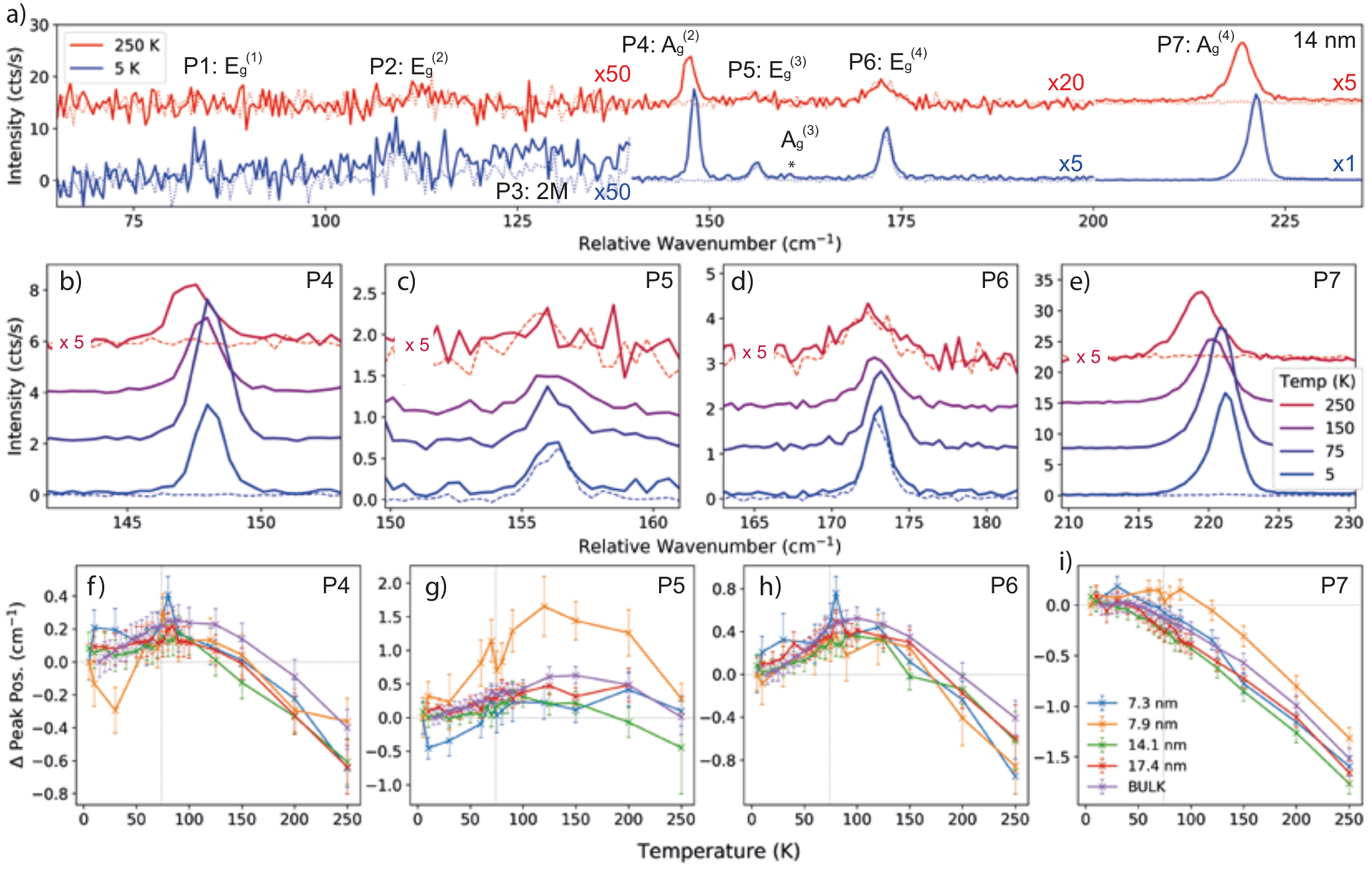}
		\caption{\label{Fig 3} {\bf Raman Spectroscopy of few layer MnPSe$_3$. a)} Raman spectra of 14 nm thin MnPSe$_3$ at low (5 K), and high (250 K), displaying 7 clearly identifiable peaks. {\bf b-e)} Temperature dependence of the phonon modes P4 through to P7. Intensity is increased for visual purposes with the given magnification factors. {\bf f-i)} Change in peak position as a function of temperature measured in a range of MnPSe$_3$ sample thicknesses from 7.3 nm (9 L) to $>$200 nm (bulk). The initial peak position is set to 0 and is highlighted with a horizontal dotted line. Peak position is seen to be mostly consistent throughout the full range of thicknesses tested. Variations seen in P5 are explored further in the main text and SI.}
	\end{figure}
	
	Moving towards thinner samples, we investigate a range of sample thicknesses down to 9 layers\supercite{Ni2021a} (7.3 nm), as determined by atomic force microscopy. As the thicknesses of the sample is decreased, we note a corresponding reduction in overall intensity of the Raman signal which can be clearly seen in the signal to noise ratio observed in Figure \ref{Fig 3}a-e. To compensate this drop in collected intensity, the acquisition exposure time is increased accordingly. The Raman signal obtained from a 14 nm (18 layer\supercite{Ni2021a}) MnPSe$_3$ sample at low (5 K) and high (250 K) temperature is shown in Figure \ref{Fig 3}a-e. Similarly to the bulk samples, a Voigt fit is applied at each temperature increment and change in peak position is extracted and displayed in Figure \ref{Fig 3}f-i. The temperature dependent Raman response is very similar to that of the bulk samples. For reference, the data for bulk sample 2 shown in Figure \ref{Fig 2} is also shown here in Figure \ref{Fig 3}f-i. It should be noted that the accuracy of fitting is reduced as the signal to noise ratio in the thinner samples is diminished, producing noisier dependencies in Figure \ref{Fig 3}f-i for the thinnest two samples (9 and 10 layers, respectively)\supercite{Ni2021a}. Even so, they generally match well with the bulk data, suggesting any out of plane inter-layer magnetic coupling component that could disrupt the antiferromagnetic ordering\supercite{Calder2021} is not significant for the range of thicknesses studied here.
	
\begin{figure}[t!]
		\centering
		\includegraphics[width=0.9\textwidth]{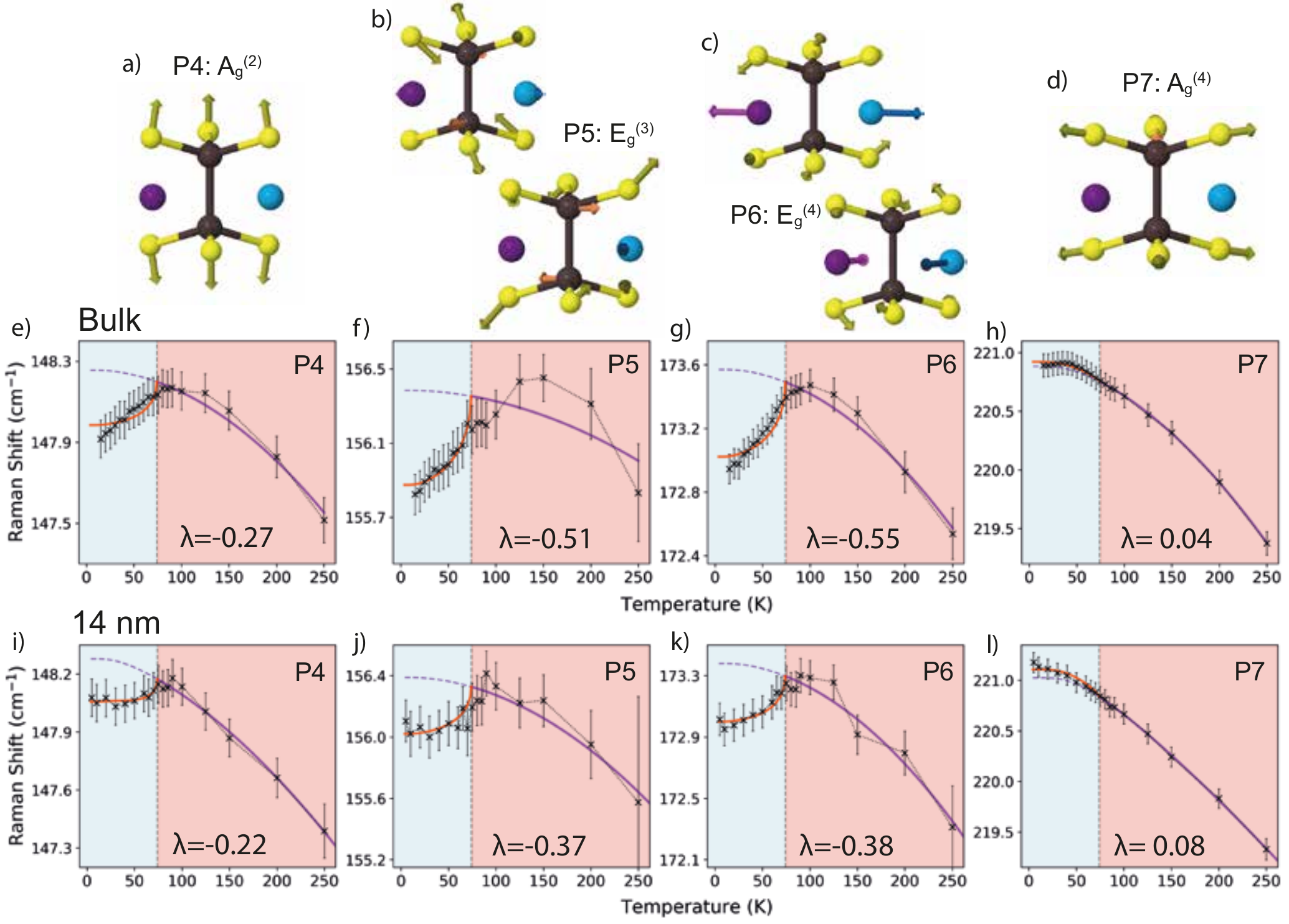}
		\caption{\label{Fig 4} {\bf Spin-phonon coupling. a-d)} Vibrational modes of MnPSe$_3$ as calculated by DFT for a) P4, b) P5, c) P6, and d) P7. Manganese atoms are shown as light and dark blue solid spheres, highlighting the alternating spin states, phosphorous atoms are brown spheres, and selenium are yellow spheres. Arrows represent the atomic shifts (from equilibrium) {\bf e-l)} Temperature dependent Raman shift of e-h) Bulk and i-l) few layer MnPSe$_3$, with emphasis on the transition from an ordered antiferromagnetic state (blue shaded region) to a disordered paramagnetic state (Red shaded region) about the N{\'e}el Temperature (dotted grey line) at 74 K. Black crosses show the experimental data with errors associated with peak fitting, while the calculated spin-phonon coupling and anharmonic phonon dependencies are shown in orange and purple solid lines, respectively. The extracted strength of the spin-phonon coupling, $\lambda$, is given for each Raman mode.}
	\end{figure}

	In order to fully analyse the temperature dependence of the Raman modes we identify a key figure of merit, the spin-phonon coupling strength, $\lambda$,\supercite{Ghosh2021,Casto2015,Fainstein2000,Prosnikov2018} which represents how strongly the magnetic spin-ordering within the Mn sub-lattice influences the atomic vibrations of the non-magnetic sub-lattice via the super-exchange pathway. In this case, $\lambda$ manifests as the difference between the observed Raman shift at T $\approx$ 0 K, and that expected by the temperature dependence without magnetic coupling (See SI Note 2 for a detailed interpretation). As such, the temperature dependent peak positions of each phonon mode, and flake thickness, are fitted using a combination of Brillouin function,\supercite{Darby1967,Wiedenmann1981a,Prosnikov2018,Ghosh2021} and anharmonic phonon model. The Brillouin function is used to define the overall magnetisation state of our 2D system as a function of temperature and is defined by the material properties. Figure \ref{Fig 4} displays the results of this combined anharmonic phonon plus Brillouin function fitting applied to bulk and few-layer (14 nm, 18 layers\supercite{Ni2021a}) MnPSe$_3$ samples, with emphasis on the transition from ordered antiferromagnetism below 74 K (pale blue regions) to the disordered paramagnetism above 74 K (red regions). The experimental data can be seen as black crosses obtained via Voigt peak fitting of each Raman mode. The error bars shown are derived from a combination of errors associated with fitting the phonon modes and spectrometer resolution. The anharmonic temperature dependence, which defines the non-magnetic temperature dependent behaviour of the phonon modes, is applied to the experimental data above the N{\'e}el temperature (solid purple line).\supercite{Balkanski1983,Lan2012,Prosnikov2018} The magnetic dependence of the phonon modes, is defined as the difference between the anharmonic fitting extended below T$_N$ (dashed purple line), and the experimentally obtained data points (black crosses). This magnetisation dependence is recreated in each phonon mode (solid orange line) using the MnPSe$_3$ reduced magnetism, defined by the Brillouin function normalised to 0 at T$_N$ and 1 at T = 0 K, multiplied by the spin-phonon coupling, $\lambda$.\supercite{Darby1967,Wiedenmann1981a,Ghosh2021,Casto2015,Fainstein2000,Prosnikov2018,Lan2012,Balkanski1983}
	
\begin{table}[ht]
	\caption{Spin-phonon coupling strength, $\lambda$, as determined by the Brillouin function and anharmonic phonon fitting procedures.}	
	\centering
	\begin{tabular}{|| c || c | c | c | c ||}
	    \hline
	    Sample Thickness & \multicolumn{4}{c ||}{Spin-phonon coupling, $\lambda$} \\
		\hline
		\hline
		Raman Mode & P4 (148 cm$^{-1}$) & P5 (156 cm$^{-1}$) & P6 (173 cm$^{-1}$) & P7 (221 cm$^{-1}$) \\
		\hline
		\hline
        7.3 nm (9 L) & -0.19 & -0.45 & -0.57 & -0.07 \\
		\hline
		7.9 nm (10 L) & -0.23 & -0.99 & -0.43 & -0.18 \\
		\hline
		14.1 nm (18 L) & -0.17 & -0.37 & -0.38 & 0.08 \\
		\hline 
		17.4 nm (22 L) & -0.17 & -0.37 & -0.42 & 0.11 \\
		\hline
		\hline
		Bulk 1 & -0.21 & -0.38 & -0.39 & 0.06 \\
		\hline
		Bulk 2 & -0.27 & -0.51 & -0.55 & 0.04 \\
		\hline
		Bulk 3 & -0.31 & -0.53 & -0.57 & 0.00 \\
		\hline
		\hline
	\end{tabular}
	\label{Table 1}
\end{table}	

	Fitting parameters of all sample thicknesses, along with a full description of the fitting process is available in SI Note 2. Table \ref{Table 1} lists the spin-phonon coupling strength, $\lambda$, of all thicknesses measured. The variation in $\lambda$, for a given phonon mode, between samples does not show any obvious trend with respect to sample thicknesses. Therefore, we attribute these discrepancies to a combination of random variations in sample strain and reduced signal-to-noise ratio in the thinner samples. It can be seen, however, that each sample measured, regardless of sample thickness, shows the same overall trends with P5 and P6 displaying the largest spin-phonon coupling strength, with P7 the lowest providing an overall consistency between all samples measured.
	
	
	Density functional theory provides extra insight into the atomic vibrational composition of the phonon modes (See Figs.\ref{Fig 4}a-d) and is used here to qualitatively analyse the relationship between magnetic spin-ordering and atomic movements via the super-exchange. A specific consideration is made to the Mn-Se-Mn ionic bond lengths and bond angle, which provides the greatest influence over the strength of the super-exchange pathway between neighbouring manganese atoms.\supercite{Vaclavkova2020,Calder2021} Full details are given in SI Note 1 but, briefly, we used both frozen phonon\supercite{Yin1982} and perturbation theory\supercite{RevModPhys.73.515} approaches within density functional theory at the generalised gradient approximation level\supercite{perdew1996generalized,PhysRevB.79.075126} including the local spin-density approximation and taking into account a possible on-site Hubbard $U$\supercite{PhysRevB.103.045141,PhysRevB.98.085127} term whose value was determined as part of this work. A frozen phonon approach calculates the inter-atomic forces with zero vibrations to construct a force constant matrix by which the normal modes at a given wave-vector can be calculated, essentially forming the phonon dispersion. Density functional perturbation theory goes a step further by introducing perturbations to the system in order to measure the response in spin and charge densities, as well as polarizability. In order to account for non-homogeneity of spin and charge densities, a generalized gradient approximation is added, while the Hubbard parameter accounts for electron localisation and hopping.\supercite{RevModPhys.73.515}
	
	Figure \ref{Fig 4}a shows that the atomic vibrational composition of P4 (148 cm$^{-1}$) is entirely due to the Se atoms shifting out-of-plane, with zero contribution from the manganese sub-lattice. This leads to the conclusion that the spin-phonon coupling, $\lambda$, observed in P4 is entirely due to the Mn-Se-Mn super-exchange pathway, with a significant variation in both pathway length and ionic bond angle leading to a large spin-phonon coupling, $\lambda$, as shown in Table \ref{Table 1}. 
	
	The atomic vibrational calculations suggest P5 (156 cm$^{-1}$) should possess a relatively low spin-phonon coupling due to the low vibrational contribution from the Mn sub-lattice. Experimental evidence however, indicates a large spin-phonon coupling, as shown in Table \ref{Table 1}. Further investigations (See SI Note 3) show a merging with a close by mode (labelled * in Figs. \ref{Fig 1}a and \ref{Fig 2}a) which has the effect of broadening the observed peak near the N{\'e}el temperature where the two separate modes are closest, and shifting the fitted peak centre to higher relative wavenumbers. When applying the temperature dependent Brillouin fitting to P5 we therefore obtain a higher experimental spin-phonon coupling from the experimental results than expected via DFT. Taking this into account with a multiple peak analysis reduces the extracted spin-phonon coupling to a value more compatible with DFT results, as explored in SI Note 3. Since this multiple peak fitting can only be applied to samples with large signal (i.e. Bulk samples with large excitation powers), the results presented in Table \ref{Table 1} are without taking this mode merging into account.
	
	P6 (173 cm$^{-1}$) is expected to show the largest spin-phonon coupling due to the large contribution from the Mn sub-lattice, along with a large change in both Mn-Se-Mn bond length and angle. This expectation is fulfilled by the experimental data displayed in Figs.\ref{Fig 4} and \ref{Table 1}. The thinner samples generally have a reduced spin-phonon coupling strength, $\lambda$, compared to the bulk samples.
	
	Finally, P7 (221 cm$^{-1}$) is observed to have a near-zero spin-phonon coupling, as shown by Figure \ref{Fig 4}. Considering the atomic shifts of this vibrational mode, it can be seen that the Mn-Se-Mn bond angle is not perturbed, while the change in bond lengths are symmetrical and small when compared to the other modes. Combined with zero magnetic sub-lattice contribution, this provides little framework for neither spin-phonon coupling through a super-exchange path, nor direct coupling.
	
	\section*{Conclusions}
	
	This work represents an investigation into an overlooked area of the opto-mechanical properties of MnPSe$_3$. The higher frequency phonon modes, predominantly attributed to the non-magnetic sub-lattice of MnPSe$_3$, have now been thoroughly explored through a combination of density functional theory and Raman spectroscopy. A previous investigation detailing the temperature dependent mode shifts of MnPSe$_3$\supercite{Liu2020c} suggested a linear dependence of peak position towards lower frequencies in both bulk and few layer MnPSe$_3$ samples. Our observations deviate significantly from that conclusion. We believe that the higher resolution grating used here, as well as pursuing a lower temperature region allows us to obtain a more accurate picture of the temperature dependent magnetic ordering than has previously been reported. Furthermore, we also note that significantly more samples are measured here, all repeating the same nonlinear temperature dependence. Our analytical approach is, likewise, more complete. We use a combination of anharmonic temperature dependent shift and magnetic specific Brillouin-like function to extract a spin-phonon coupling strength, $\lambda$, from each phonon resonance. 
 

	\section*{Methods}
	\noindent	
	{\bf Temperature dependent Raman spectroscopy.}
	Ultra-low frequency Raman spectroscopy is performed using a custom-built free-space micro-Raman setup. Flakes of MnPSe$_3$ are placed within a flow cryostat (MicrostatHiRes, Oxford Instruments) and cooled to $\sim 5$ K. A flow controller and heater is used to control the sample temperature as required. The phonon modes of MnPSe$_3$ are excited using a 532 nm diode pumped solid state laser (model 04-01, Cobalt) with linewidth $<$ 1 MHz. The excitation beam is focused onto the samples via a high (0.55) NA 50x microscope objective (M Plan Apo 50X, Mitutoyo). Back-scattered light is collected using the same objective lens. The laser line is rejected using three 10 cm$^{-1}$ Bragg notch filters (BragGrate\textsuperscript{\texttrademark}, Optigrate). The Raman signal is analysed using a 0.75 cm monochromater (SP750, Princeton Instruments) with a 1800 g/mm holographic grating and a nitrogen cooled charge-coupled device camera with a pixel size of 20 $\mu$m x 20 $\mu$m (PyLon:100BR, Princeton Instruments) providing spectral resolution of $\approx 0.4$ cm$^{-1}$ per pixel. The linewidths of the phonons studied here are larger than one pixel, allowing for sub-pixel precision when finding peak position, see SI Note 2. Polarisation optics are configured as, quarter wave plate - motorised linear polariser - Sample - motorised linear polariser - motorised half wave plate, to allow for linear polarisation excitation of arbitrary polarisation angle without loss of power, selection of co- and cross- polarisation detection without loss of signal to the holographic grating. Laser power at the sample is typically 0.5 mW (Power density of 6.37$\times10^{4}$ W/cm$^{2}$) unless stated otherwise. We observe no effect from laser heating in our data (No shift in N\'eel temperature towards lower temperatures), thanks in part to the large cooling power of the flow cryostat. The spectral resolution of this system is $\approx$ 0.4 cm$^{-1}$, with spatial resolution of $\approx$ 1 $\mu$m. A schematic of the experimental setup is available in SI Note 7.
	
	\noindent
	{\bf Density Functional Theory.} We used both frozen phonon and perturbation theory approaches within density functional theory at the generalised gradient approximation level of approximation including the local spin-density approximation and taking into account a possible on-site Hubbard $U$ term whose value was determined as part of this work. A full description of the DFT analysis is provided in SI Note 1.	
	
	
	\section*{Acknowledgments}
	We acknowledge IT services at the Universities of Sheffield and Bath for the provision of services for High Performance Computing (HPC). Computational work in Bath was supported by the EU Horizon 2020 OCRE project ``Cloud funding for research''. D. J. G., O. H., and A. I. T. acknowledge support from the European Graphene Flagship Project under grant agreement number 881603 and EPSRC grants EP/S030751/1, EP/V006975/1, and EP/V026496/1.

	\subsection*{Author Contributions}
	D.J.G. and O.H. carried out micro-mechanical exfoliation of the MnPSe$_3$ flakes and performed experimental observations using Raman spectroscopy. D.W. performed density functional theory of MnPSe$_3$. D.W. provided insight into the analytical model used to fit the temperature dependent Raman response while D.J.G. performed the fitting procedures. D.J.G., D.W., O.M., and A.I.T. had ongoing analytical discussions regarding the experimental and DFT data throughout the project. D.J.G., D.W., and A.I.T. managed various aspects of the project. A.I.T. conceived of, and supervised the project. D.J.G. wrote the manuscript with major contributions from DW and further contributions from all co-authors.
	
	\subsection*{Data Availability}
	\noindent
	The data that support the plots within this paper and other findings of this study are available from the corresponding author upon reasonable request.
	
	\subsection*{Code availability}
	The codes that support the findings of this study are available from \\https://www.quantum-espresso.org/. Inputs to these codes are available free of charge at https://doi.org/10.15125/BATH-01218

	\subsection*{Competing interests}
	\noindent
	The authors declare no competing interests.
	
	
	\def\bibsection{\section*{References}}
	\printbibliography{}
	
\end{document}